\documentclass[letter,twocolumn]{jpsj2} 
\renewcommand{\Vec}[1]{\mbox{\boldmath$#1$}}
\title{The effect of interchain interaction on the 
pairing symmetry competition in organic superconductors (TMTSF)$_2$X}

\author{Kazuhiko \textsc{Kuroki}$^{1}$ and Yukio \textsc{Tanaka}$^{2}$}

\inst{$^{1}$Department of Applied Physics and
Chemistry, The University of Electro-Communications,
Chofu, Tokyo 182-8585, Japan\\
$^{2}$ Department of Applied Physics,
Nagoya University, Nagoya, 464-8603, Japan\\
}

\abst{We investigate the effect of interchain repulsive interaction 
on the pairing symmetry competition in quasi-one-dimensional 
organic superconductors (TMTSF)$_2$X by applying random 
phase approximation and quantum Monte Carlo calculation to an 
extended Hubbard model.
We find that interchain repulsive interaction enhances the 
$2k_F$ charge fluctuations, thereby making the possibility 
of spin-triplet $f$-wave pairing dominating over spin-singlet $d$-wave 
pairing realistic.}

\kword{(TMTSF)$_2$X, superconductivity, spin-triplet pairing, 
interchain interaction, charge and spin fluctuations}

\begin{document}
\maketitle

\begin{figure}[tb]
\begin{center}
\includegraphics[width=8cm,clip]{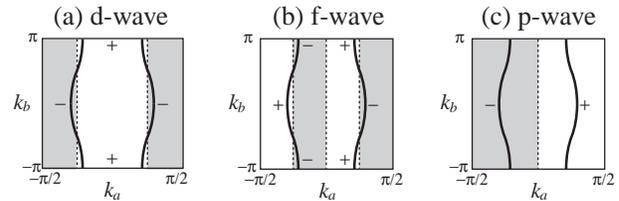}
\end{center}
\caption{Candidates for the gap function of (TMTSF)$_2$X 
are schematically shown along with the Fermi surface (solid curves). 
The dashed lines represent the nodes of the gap, 
whose $k_b$ dependence is omitted
for simplicity. (For the actual $k_b$ dependence, see 
Fig.\protect\ref{fig3}.) We call 
the gap in fig.(a)((b)) $d$-wave ($f$-wave) in the sense that the gap 
changes sign as $+-+-$ ($+-+-+-$) along the Fermi surface.}
\label{fig1}
\end{figure}

Possible occurrence of unconventional superconductivity 
in organic conductors has been of great interest recently.
Microscopically understanding 
the mechanism of pairing in those materials is 
an intriguing theoretical challenge.
Among the various candidates of unconventional superconductors, 
the pairing mechanism  of 
quasi-one-dimensional (q1D) organic superconductors 
$\mbox{(TMTSF)}_{2}X$
($X=\mbox{PF}_{6}$, $\mbox{ClO}_{4}$, etc.),
so called the Bechgaard salts,\cite{Jerome,Bechgaard}
has been quite puzzling.
Namely, since superconductivity lies right next to the $2k_{\rm F}$ 
spin density wave (SDW) phase 
in the pressure-temperature phase diagram,
a spin-singlet $d$-wave-like pairing (shown schematically in 
Fig.\ref{fig1}(a)) is expected to take place 
as suggested by several authors.\cite{Shima01,KA99,KK99}.
However, 
an unchanged Knight shift across $T_c$ \cite{Lee02} 
and a large $H_{c2}$ exceeding the Pauli limit\cite{Lee00} 
suggest a realization of spin-triplet pairing.
As for the orbital part of the order parameter, 
there have been NMR experiments 
suggesting the existence of nodes and thus unconventional 
pairing,\cite{Takigawa} 
although a thermal conductivity measurement suggests absence of 
nodes for (TMTSF)$_2$ClO$_4$.\cite{BB97}
As a possible solution for this puzzle of spin-triplet pairing,
one of the present authors 
has phenomelogically proposed 
that triplet $f$-wave-like 
pairing (whose gap is shown schematically in Fig.\ref{fig1}(b)) 
may take place 
due to a combination of quasi-1D (disconnected) Fermi surface 
and the coexistence of $2k_{\rm F}$ spin 
and $2k_{\rm F}$ charge fluctuations.\cite{KAA01} 
Namely, due to the disconnectivity of the Fermi surface,
the number of gap nodes that intersect the Fermi surface is the 
same between $d$ and $f$. Moreover, if the $2k_F$ spin 
and charge fluctuations have about the same magnitude, 
spin-singlet and spin-triplet pairing interactions have 
close absolute values (with opposite signs) as will be explained later. 
In such a case, spin-triplet $f$-wave pairing should be 
closely competitive against singlet $d$-wave pairing.
As for other possibilities of triplet pairing, 
the $p$-wave state in which
the nodes of the gap (Fig.\ref{fig1}(c)) do not intersect
the Fermi surface has been considered 
from the early days,\cite{Abrikosov,HF87,Lebed} 
but from a microscopic 
point of view, spin-triplet pairing interaction has a negative 
sign for the momentum transfer of $2k_F$ unless spin fluctuations are 
highly anisotropic, so that a gap that changes sign between the left and 
right portions of the Fermi surface is unlikely to take place.
\cite{Kohmoto}
A similar phenomelogical proposal of $f$-wave pairing 
in (TMTSF)$_2$X has also been given by Fuseya {\it et al.}\cite{Fuseya1}
Experimentally, the $f$-wave scenario due to the coexistence of 
$2k_F$ spin and charge fluctuations is indirectly supported by the 
observation that $2k_F$ charge density wave (CDW) actually coexists 
with $2k_F$ SDW in the insulating phase lying next to the 
superconducting phase.\cite{Pouget,Kagoshima}  

As for {\it microscopic} theories for the pairing competition,
we have previously shown using a ground state quantum Monte Carlo method that 
$f$-wave strongly dominates over $p$-wave in the Hubbard model 
that considers only the on-site repulsive interaction.\cite{KTKA}
More recently, we have shown, by applying random phase 
approximation (RPA) to an extended Hubbard model,   
that $f$-wave pairing can indeed dominate over $d$-wave pairing 
when we have large enough 
second nearest neighbor repulsion $V'$,\cite{TanakaKuroki04} which has been 
known for some years to have the effect of stabilizing
$2k_F$ CDW configuration.\cite{Kobayashi,Suzumura}
To be more precise, the condition for $f$-wave dominating over 
$d$-wave is to have $V'\simeq U/2$ (where $U$ is the on-site repulsion) 
or larger $V'$ because 
$2k_F$ spin and $2k_F$ charge fluctuations have the same magnitude for 
$V'=U/2$ within RPA.
A similar condition for $f$-wave being competitive against $d$-wave 
has also been obtained in a recent renormalization group study.\cite{Fuseya04}
Although these results do suggest that $f$-wave pairing can indeed 
be realized in microscopic models, the condition 
that the {\it second} nearest neighbor repulsion being nearly equal to or 
larger than half the 
{\it on-site} repulsion may not be realized so easily in actual materials. 
In the present study, 
we consider a model where the {\it interchain} 
repulsion is taken into account, which turns out to give a more 
realizable condition for $f$-wave dominating over $d$-wave  
due to the enhancement of $2k_F$ charge fluctuations.
After completing the major part of this study, we came to notice that 
a similar conclusion has been reached quite recently 
using a renormalization group approach.\cite{Nickel}

\begin{figure}[tb]
\begin{center}
\includegraphics[width=8cm,clip]{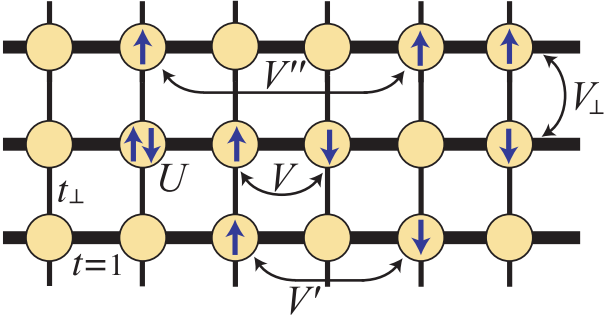}
\end{center}
\caption{The model of the present study is shown.}
\label{fig2}
\end{figure}

The model considered in the present study 
is shown in Fig.\ref{fig2}. In standard notations,
the Hamiltonian is given as 
\[
H=-\sum_{<i,j>,\sigma} 
t_{ij}c^{\dagger}_{i\sigma}c_{j\sigma}
+U\sum_{i}n_{i\uparrow}n_{i\downarrow}
+ \sum_{<i,j>}V_{ij} n_{i}n_{j},
\]
where $c^{\dagger}_{i\sigma}$ creates a hole (note
that (TMTSF)$_2$X is actually a 3/4 filling system in the electron picture) 
with spin $\sigma = \uparrow, \downarrow$ at site $i$.
As for the kinetic energy terms, we consider nearest neighbor hoppings  
$t_{ij}=t$ in the (most conductive) $a$-direction 
and $t_{ij}=t_\perp$ in the $b$-direction. 
$t$ is taken as the unit of energy, 
and we adopt $t_\perp=0.2t$ throughout the study.
$U$ and $V_{ij}$ are the on-site and the off-site repulsive interactions,
respectively, where 
we take into account the nearest neighbor {\it interchain}
repulsion $V_\perp$ in addition to the intrachain 
on-site ($U$), nearest ($V$), next nearest ($V'$), and third 
nearest ($V''$) neighbor repulsions 
considered in our previous study.\cite{TanakaKuroki04}

%
%
Within RPA\cite{Scalapino,TYO,KTOS}, 
the effective pairing interactions for the singlet and 
triplet channels due to spin and charge fluctuations are given as 
\begin{align}
\label{1}
V^{s}(\Vec{q})=
U + V({\Vec q}) + \frac{3}{2}U^{2}\chi_{s}(\Vec{q})
\nonumber\\
-\frac{1}{2}(U + 2V({\Vec q}) )^{2}\chi_{c}(\Vec{q})
\end{align}
\begin{align}
\label{2}
V^{t}(\Vec{q})=
V({\Vec q}) - \frac{1}{2}U^{2}\chi_{s}(\Vec{q})
\nonumber\\
-\frac{1}{2}(U + 2V({\Vec q}) )^{2}\chi_{c}(\Vec{q}),
\end{align}
where 
\begin{equation}
V(\Vec{q})=2V\cos q_{x} + 2V'\cos(2q_{x}) + 2V''\cos(3q_{x})
+2V_\perp\cos(q_y)
\label{3}
\end{equation}
Here, $\chi_{s}$ and $\chi_{c}$ are the spin and charge 
susceptibilities, respectively,  which are given as 
\begin{align}
\label{4}
\chi_{s}(\Vec{q})=\frac{\chi_{0}(\Vec{q})}
{1 - U\chi_{0}(\Vec{q})}
\nonumber\\
\chi_{c}(\Vec{q})=\frac{\chi_{0}(\Vec{q})}
{1 + (U + 2V(\Vec{q}) )\chi_{0}(\Vec{q})}.
\end{align}
Here $\chi_{0}$ is the bare susceptibility given by 
\[
\chi_{0}(\Vec{q})
=\frac{1}{N}\sum_{\Vec{p}} 
\frac{ f(\epsilon_{\Vec{p +q}})-f(\epsilon_{\Vec{p}}) }
{\epsilon_{\Vec{p}} -\epsilon_{\Vec{p+q}}}
\]
with
$\epsilon_{\Vec{k}}=-2t\cos k_a -2t_\perp\cos k_b - \mu$ and 
$f(\epsilon_{\Vec{p}})=1/(\exp(\epsilon_{\Vec{p}}/T) + 1)$. 
$\chi_0$ peaks at the nesting vector 
$\Vec{Q}_{2k_F}$ ($=(\pi/2,\pi)$ here) of the Fermi surface. 
%
%
%

To obtain $T_c$, we solve the linearized gap equation within the 
weak-coupling theory, 
\begin{equation}
\lambda^{s,t} \Delta^{s,t}(\Vec{k})
=-\sum_{\Vec{k'}} V^{s,t}(\Vec{k-k'})
\frac{ \rm{tanh}(\beta \epsilon_{{\Vec{k'} }}/2) }{2 \epsilon_{\Vec{k'}} }
\Delta^{s,t}(\Vec{k'}). 
\end{equation}
The eigenfunction $\Delta^{s,t}$ of this eigenvalue equation is the 
gap function. 
The transition temperature $T_c$ is determined as the temperature 
where the eigenvalue $\lambda$ reaches unity. Note that 
the main contribution to the summation in the right hand side comes from 
$\Vec{k-k'}\simeq\Vec{Q}_{2k_F}$ because $V^{s,t}(\Vec{q})$ peaks around 
$\Vec{q}=\Vec{Q}_{2k_F}$.
Although RPA is quantitatively insufficient for discussing the 
absolute values of $T_c$, 
we expect this approach to be valid for studying 
the {\it competition} between different pairing symmetries. 

Now, from eqs.(\ref{3}) and (\ref{4}), 
it can be seen that $\chi_c(\Vec{Q}_{2k_F})=
\chi_s(\Vec{Q}_{2k_F})$ holds when $V'+V_{\perp}=U/2$, which 
in the absence $V_\perp$ of course 
reduces to the condition $V'=U/2$ obtained in our previous study. 
This in turn results in $V^s(\Vec{Q}_{2k_F})=-V^t(\Vec{Q}_{2k_F})$ 
for the pairing interactions apart from the first order terms 
as can be seen from eqs.(\ref{1}) and (\ref{2}).   
Thus, considering the fact that the number of nodes intersecting the 
Fermi surface is the same between $d$ and $f$,  
the condition for $f$-wave being competitive against $d$-wave should be 
$V'+V_{\perp}\simeq U/2$.
The possibility of this condition being satisfied in actual materials 
is realistic since 
$V_{\perp}$ can be comparable with the intrachain off-site repulsions 
due to the fact that 
the lattice constant in the $a-$ and $b-$ directions are of the 
same order. An intuitive picture here is that $V_\perp$ tends to 
``lock'' more firmly the $2k_F$ charge configuration induced by $V'$, 
so that $2k_F$ charge fluctuations 
are enhanced, thereby stabilizing the spin-triplet $f$-wave state.

\begin{figure}[tb]
\begin{center}
\includegraphics[width=8cm,clip]{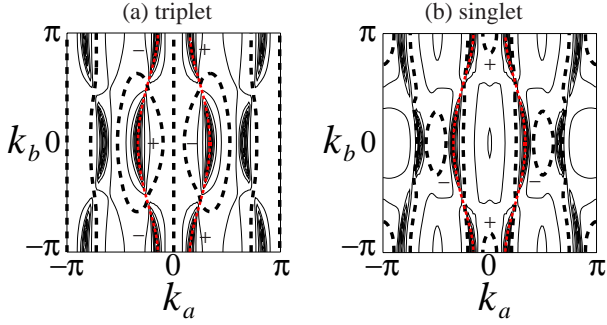}
\end{center}
\caption{The gap functions having the largest 
eigenvalue for the (a)triplet and (b)singlet pairing channels.
The parameter values are taken as $U=1.7$, $V=0.8$, $V'=0.45$, 
$V''=0.2$, $V_\perp=0.4$ and $T=0.011$ (=$T_c$ of the $f$-wave pairing). 
The dark dashed curves represent the nodes of the gap, while 
a pair of light dotted curves near $k_a=\pm\pi/4$ is 
the Fermi surface.}
\label{fig3}
\end{figure}

\begin{figure}[tb]
\begin{center}
\includegraphics[width=8cm,clip]{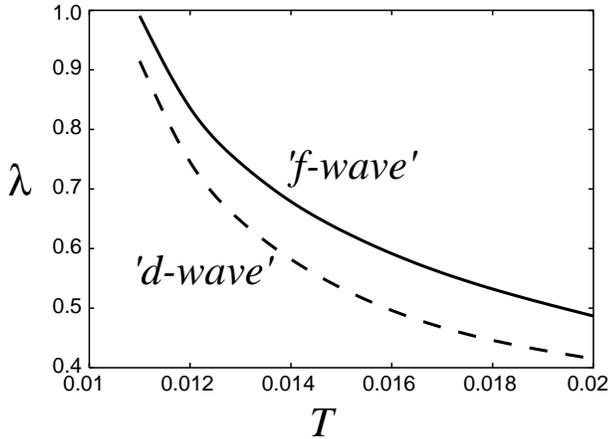}
\end{center}
\caption{The largest eigenvalue in the singlet and the triplet 
channels are plotted as functions of temperature for 
$U=1.7$, $V=0.8$, $V'=0.45$, $V''=0.2$, and $V_\perp=0.4$.}
\label{fig4}
\end{figure}

Bearing the above analysis in mind, 
we now move on to the RPA calculation results for the pairing symmetry 
competition between $f$- and $d$-waves.
We first focus on the case where the parameter values satisfy the 
condition for $\chi_c(\Vec{Q}_{2k_F})=\chi_s(\Vec{Q}_{2k_F})$, that is 
when $V'+V_\perp=U/2$ holds. Here we take $U=1.7$, $V=0.8$, $V'=0.45$,
$V''=0.2$, and $V_\perp=0.4$ in units of $t$. 
Note that $V'$ is much smaller than $U/2$.
As expected, 
the singlet pairing having the largest eigenvalue $\lambda$ has a 
$d$-wave gap, while the triplet pairing with the largest $\lambda$ has 
a $f$-wave gap, as seen in Fig.\ref{fig3}.
In Fig.\ref{fig4}, we plot $\lambda$ as functions of temperature 
for $d$-wave and $f$-wave pairings.
The two pairings closely compete with each other, 
but $f$-wave pairing dominates over $d$-wave pairing and gives a 
higher $T_c$. $f$-wave not being degerate with $d$-wave
even for $V'+V_\perp=U/2$ is due to the effect of the 
first order terms in eqs.(\ref{1}) and (\ref{2}) 
as discussed in our previous study.\cite{TanakaKuroki04}

\begin{figure}[tb]
\begin{center}
\includegraphics[width=8cm,clip]{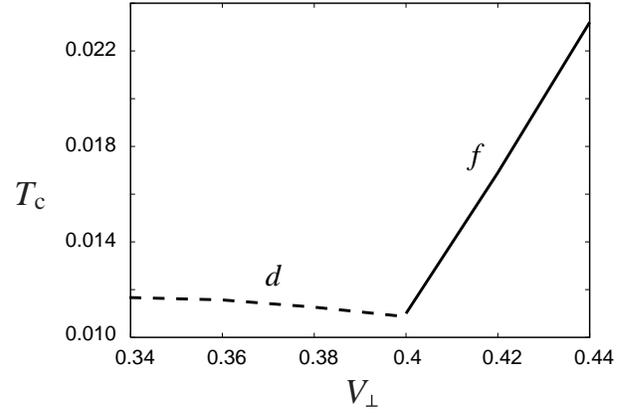}
\end{center}
\caption{$T_c$ plotted as a function of $V_\perp$ for  
$U=1.7$, $V=0.8$, $V'=0.45$, and $V''=0.2$. The solid (dashed) 
curve represent the $f$ ($d$)-wave regime.}
\label{fig5}
\end{figure}

To look into the effect of $V_\perp$ on the $f$- vs. $d$- competition 
in more detail, we plot $T_c$ along with the pairing symmetry 
as a function of $V_\perp$ in Fig.\ref{fig5}.
The pairing symmetry is $f$-wave  and $T_c$ increases 
with $V_\perp$ for $V_\perp \geq U/2-V'$(=0.4 here),  
while the pairing occurs in the $d$-wave channel 
with a nearly constant $T_c$ for $V_\perp < U/2-V'$. 
The increase of the $f$-wave $T_c$ 
is due to the enhancement of $2k_F$ charge fluctuations with 
increasing $V_\perp$.

\begin{figure}[tb]
\begin{center}
\includegraphics[width=8cm,clip]{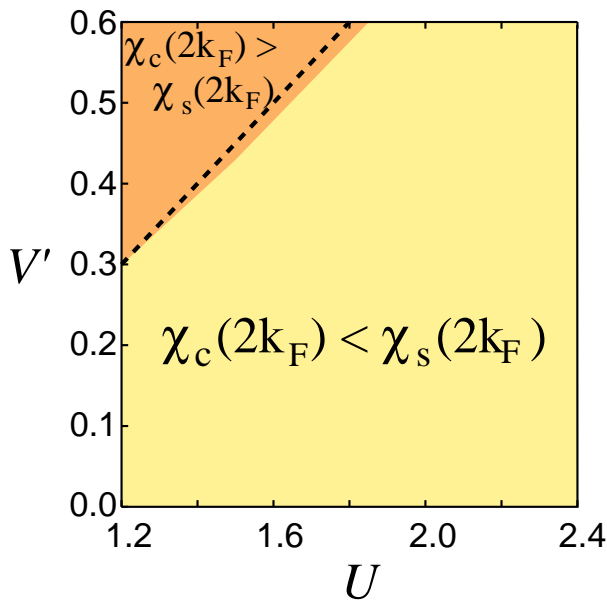}
\end{center}
\caption{AFQMC result for the competition between 
$\chi_s(\Vec{Q}_{\rm 2k_F})$ and $\chi_c(\Vec{Q}_{\rm 2k_F})$
shown in $U-V'$ space. $V=0.9$, $V''=0$, $V_\perp=0.3$, and $T=0.25$. 
The dashed line represents the 
RPA condition for $\chi_s(\Vec{Q}_{\rm 2k_F})=\chi_c(\Vec{Q}_{\rm 2k_F})$.}
\label{fig6}
\end{figure}

Finally, in order to check the validity of RPA, 
we have performed  auxiliary field quantum Monte Carlo (AFQMC) 
calculation\cite{Hirsch,ZC,White}  for the same extended Hubbard model.
Let us first briefly summarize this method. 
In AFQMC, the density operator is decomposed into 
the kinetic energy part and the interaction part using  
Trotter-Suzuki decomposition,\cite{Trotter,Suzuki} and we perform 
discrete Hubbard-Stratonovich transformation\cite{Hirsch} 
for the interaction part.
The summation over the Stratonovich variables are taken by 
Monte-Carlo importance sampling.
Using this method, correlation functions and susceptibilities 
can be calculated for finite size systems (16 sites in the $a$-direction and 
4 sites in the $b$-direction=64 sites in the present study), 
and the results are exact 
within the statistical errors. A defect of this 
approach is that we cannot go down to very low temperatures 
in the presence of off-site repulsions such as $V$, $V'$ and 
$V_\perp$ due to the negative sign problem, so that 
it is difficult to look into the pairing symmetry competition itself.
 Nevertheless, we can check the validity of 
RPA at moderate temperatures of the order of $0.1t$.
Here we compare the values of 
$\chi_s(\Vec{Q}_{2k_F})$ and $\chi_c(\Vec{Q}_{2k_F})$ calculated 
by AFQMC at $T=0.25$,fixing $V=0.9$, $V''=0$ and $V_\perp=0.3$. 
In Fig.\ref{fig6}, we show the ``phase diagram'' in $U-V'$ plane, 
where we find that the 
AFQMC boundary for 
$\chi_c(\Vec{Q}_{2k_F})=\chi_s(\Vec{Q}_{2k_F})$ 
is very close to the RPA boundary  $V'+V_\perp=U/2$.
This result suggests that the RPA condition for 
$\chi_c(\Vec{Q}_{2k_F})=\chi_s(\Vec{Q}_{2k_F})$ 
is reliable at least at 
moderate temperatures.

To summarize, we have studied the pairing symmetry competition in 
a model for (TMTSF)$_2$X which considers not only the 
intrachain repulsions but also the interchain repulsion.
We find that the possibility of satisfying the condition for 
realizing $f$-wave pairing becomes more realistic 
in the presence of the interchain repulsion.
It would be an interesting 
future study to investigate whether this condition 
is actually satisfied in (TMTSF)$_2$X using
first principles or quantum chemical calculations.
Experimentally, it would be interesting to further confirm 
spin-triplet pairing by using probes 
complementary to those in the previous studies\cite{Lee02,Lee00},
for example, a phase sensitive
tunneling spectroscopy study\cite{TK95}  
with\cite{Tanuma2} or without\cite{Tanuma1,Sengupta} 
applying a magnetic field, or those based on a newly 
developed theory for triplet superconductors, which has been  
proposed by one of the present authors.\cite{TanakaKas}

K.K. acknowledges H. Fukuyama, H. Seo, and A. Kobayashi for 
motivating us to study the effect of interchain repulsion.
He also thanks J. Suzumura and Y. Fuseya for valuable discussion.
Part of the numerical calculation has been performed
at the facilities of the Supercomputer Center,
Institute for Solid State Physics,
University of Tokyo.

\end{document}